\documentclass[thmsa]{article}

\input tcilatex

\begin{document}

\author{Babur M. Mirza\thanks{
E-mail: bmmirza2002@yahoo.com} \\
Department of Mathematics, Quaid-i-Azam University, \\
Islamabad. Pakistan. 45320}
\title{Charged Particle Dynamics in the Field of a Slowly Rotating Compact Star}
\date{August 31, 2004}
\maketitle

\begin{abstract}
We study the dynamics of a charged particle in the field of a slowly
rotating compact star in the gravitoelectromagnetic approximation to the
geodesic equation . The star is assumed to be surrounded by an ideal, highly
conducting plasma (taken as a magnetohydrodynamic fluid) with a stationary,
axially symmetric electromagnetic field. The general relativistic Maxwell
equations are solved to obtain the effects of the background spacetime on
the electromagnetic field in the linearized Kerr spacetime. The equations of
motion are then set up and solved numerically to incorporate the
gravitational as well as the electromagnetic effects. The analysis shows
that in the slow rotation approximation the frame dragging effects on the
electromagnetic field are absent. However the particle is directly effected
by the rotating gravitational source such that close to the star the
gravitational and electromagnetic field produce contrary effects on the
particle trajectories.
\end{abstract}

\section{Introduction}

The study of accretion dynamics of charged particles in the vicinity of a
compact gravitational source is of observational as well as theoretical
interest; particularly for the information it reveals about the nature and
influence of the background spacetime on the various physical processes
occurring in the star's vicinity. Whereas a compact star is formed largely
of a degenerate plasma$^{1}$, matter surrounding the star exists in a highly
ionized, though\ a less dense, state$^{2}$. In this region the charged
particles are more free to move while not accreted to relativistic
velocities, and the dynamics of these charged particles is governed not only
by the gravitational but also by the electromagnetic field of the star.
However the electromagnetic field is itself\ effected by the geometry of the
background spacetime. Therefore it is of importance and of astrophysical
relevance that the two-fold effects; namely, the coupling of the
gravitational and the electromagnetic field of the star, as well as the
direct effects of the background spacetime on the motion of a charged
particle, must be addressed together$^{3,4}$.

Here we present an investigation of these effects on the dynamics of an
accreted\ charged particle lying in the star's plasma atmosphere. To pose
the problem quantitatively we consider a slowly rotating isolated compact
star as the source of gravitational field which defines the geometry of the
background spacetime. An axially symmetric, highly conducting plasma with a
stationary, axially symmetric electromagnetic field is assumed to constitute
the star's atmosphere. We regard the electromagnetic field as perturbed by
the background spacetime, however the back-reaction of the electromagnetic
field on the spacetime curvature is neglected and the standard Kerr metric
(in its linearized form) is used for the fixed background spacetime. For the
plasma surrounding the star the magnetohydrodynamic (MHD) approximation is
assumed, as such the plasma is considered as an ideal highly conducting MHD
fluid$^{5-7}$. The formulation of the electromagnetic field equations is
based on a generalized definition of the electromagnetic field tensor for an
ideal MHD fluid in a curved spacetime$^{8,9}$. The direct effects of the
spacetime dragging on the particle dynamics are discussed in the
gravitoelectromagnetic (GEM) approximation to the geodesic equation$^{10-12}$%
. The GEM approximation, which assumes the charged particle motion to be
slow, is generally valid in the case of stable compact stars (i.e.
non-collapsing objects). Also for the case of a magnetized compact star with
a dipole field the GEM approximation is applicable except at or very close
to the poles where particle velocities can be relativistic due to intense
magnetic field. Generally the accretion occurs well below the relativistic
limit and so the GEM approximation can be used. Throughout we take the
velocity four vector as defined by a comoving observer ZAMO (zero angular
momentum observer). The ZAMO is a useful class of observers, relevant to
astrophysics, circling the gravitational source with a given angular
velocity at a fixed radial distance and polar angle$^{13,14}$.

The paper is organized as follows. In the first part of our analysis,
comprising section II, we obtain exact solutions of the general relativistic
Maxwell equations using the generalized definition of the electromagnetic
field tensor for an ideal, highly conducting MHD fluid in linearized Kerr
spacetime. The solution obtained shows that in the slow rotation
approximation the effects of rotation on the electromagnetic field are
absent. Then, in section III, assuming the star to be a sphere of
homogeneous mass density, we study numerically the dynamics of the charged
particle using the GEM approximation. In this approximation the direct
effects of rotation (i.e. via frame-dragging) on particle trajectories are
found to oppose that of the electromagnetic field, hence the particle motion
is a result of balancing and counter-balancing of these effects. Lastly, in
section IV, we give a discussion and a summary of the main conclusions of
the paper.

\section{The Electromagnetic Field}

\subsection{\protect\bigskip General Relativistic Maxwell Equations}

The exterior spacetime to a slowly rotating compact gravitational source of
mass $M$ is described by the linearized Kerr metric. In the Boyer-Lindquist
coordinates the metric can be written as the general line element:

\begin{equation}
ds^{2}=-e^{2\Phi (r)}dt^{2}-2\omega (r)r^{2}\sin ^{2}\theta dtd\varphi
+r^{2}\sin ^{2}\theta d\varphi ^{2}+e^{-2\Phi (r)}dr^{2}+r^{2}d\theta ^{2},\ 
\tag{1}
\end{equation}
where

\begin{equation}
e^{2\Phi (r)}=(1-\frac{2M}{r}),  \tag{2}
\end{equation}
and 
\begin{equation}
\omega (r)\equiv \frac{d\varphi }{dt}=-\frac{g_{t\varphi }}{g_{\varphi
\varphi }},  \tag{3}
\end{equation}
is the angular velocity of a free falling frame brought into rotation by the
frame dragging of the spacetime. Here and in what follows the Greek indices
run as $t,r,\theta ,$ and $\varphi $ respectively. Also throughout we assume
the gravitational units in which $G=1=c$.

The general relativistic form of the Maxwell equations is $\qquad $%
\begin{equation}
F_{\alpha \beta ,\gamma }+F_{\beta \gamma ,\alpha }+F_{\gamma \alpha ,\beta
}=0,  \tag{4}
\end{equation}

\begin{equation}
\left( \sqrt{-g}F^{\alpha \beta }\right) _{,\beta }=4\pi \sqrt{-g}J^{\alpha
},  \tag{5}
\end{equation}
where $g$ represents the determinant of the metric tensor $g_{\alpha \beta }$
given by the Einstein field equations. Here $F_{\alpha \beta }$ is the
generalized electromagnetic field tensor for an ideal MHD fluid given by a
unique tensorial expression:

\begin{equation}
F_{\alpha \beta }=u_{\alpha }E_{\beta }-u_{\beta }E_{\alpha }+\eta _{\alpha
\beta \gamma \delta }u^{\gamma }B^{\delta },  \tag{6}
\end{equation}
and $J^{\alpha }$ is current four vector. In general the current four vector
is the sum of two terms corresponding to a convection and to a conduction
current:

\begin{equation}
J^{\alpha }=\epsilon u^{\alpha }+\sigma u_{\beta }F^{\beta \alpha }  \tag{7}
\end{equation}
where $\epsilon $ is the proper charge density, $\sigma $ is the
conductivity of the fluid, and $u^{\alpha }$ is the unit velocity $4$%
-vector. The volume element 4-form $\eta _{\alpha \beta \gamma \delta }$ and
its dual $\eta ^{\alpha \beta \gamma \delta }$ are defined by

\begin{equation}
\eta _{\alpha \beta \gamma \delta }=\sqrt{-g}\epsilon _{\alpha \beta \gamma
\delta },\quad \eta ^{\alpha \beta \gamma \delta }=-\frac{1}{\sqrt{-g}}%
\epsilon ^{\alpha \beta \gamma \delta },  \tag{8}
\end{equation}
where $\epsilon _{\alpha \beta \gamma \delta }$ is the Levi -Civita symbol,
which is $+1,-1,$ or $0$ for an even, odd, or non-permutation of $\alpha
\beta \gamma \delta $ respectively. Also the four vectors $E_{\alpha }$ and $%
B_{\alpha }$, denote the electric and magnetic field components in the four
dimensional spacetime. For a ZAMO $u_{r}$ and $u_{\theta \text{ }}$ vanish,
and using $u^{\alpha }u_{\alpha }=-1$, the components of the four velocity
vector are:

\begin{equation}
u^{\alpha }=e^{-\Phi (r)}(1,0,0,\omega (r)),\quad u_{\alpha }=e^{\Phi
(r)}(-1,0,0,0).  \tag{9}
\end{equation}

Here we employ the often made assumption of a perfectly conducting plasma
surrounding the star (cf e.g. [5], [14] for a discussion of this assumption
with respect to the generalized Ohm's law). Since for plasma the condition
of neutrality also holds thus $J^{\alpha }$ must vanish identically. The
Maxwell equations (4) give:

\begin{equation}
(\sqrt{-g}u^{t}B^{r})_{,r}+(\sqrt{-g}u^{t}B^{\theta })_{,\theta }=0,\qquad 
\tag{10}
\end{equation}
\begin{equation}
(u_{t}E_{\varphi })_{,r}=0,  \tag{11}
\end{equation}
\begin{equation}
(u_{t}E_{\varphi })_{,\theta }=0,  \tag{12}
\end{equation}
\begin{equation}
(u_{t}E_{\theta }-\sqrt{-g}u^{\varphi }B^{r})_{,r}-(u_{t}E_{r}-\sqrt{-g}%
u^{\varphi }B^{\theta })_{,\theta }=0,  \tag{13}
\end{equation}
whereas from expression (5) we obtain:

\begin{equation}
(\sqrt{-g}u^{t}E^{r})_{,r}+(\sqrt{-g}u^{t}E^{\theta })_{,\theta }=0, 
\tag{14}
\end{equation}
\begin{equation}
(u_{t}B_{\varphi })_{,r}=0,  \tag{15}
\end{equation}
\begin{equation}
(u_{t}B_{\varphi })_{,\theta }=0,  \tag{16}
\end{equation}
\begin{equation}
(u_{t}B_{\theta }-\sqrt{-g}u^{\varphi }E^{r})_{,r}-(u_{t}B_{r}-\sqrt{-g}%
u^{\varphi }E^{\theta })_{,\theta }=0.  \tag{17}
\end{equation}

\subsection{\protect\bigskip Solution to the Maxwell Equations}

The electromagnetic field outside the plasma surrounding the compact star is
now determined by equation (10) to (17). To solve this system of equations
let us assume the following separation ansatz for the magnetic field
components $B^{r}$ and $B^{\theta }$:

\begin{equation}
B^{r}(r,\theta )=R_{B}^{r}(r)\Theta _{B}^{r}(\theta ),\quad B^{\theta
}(r,\theta )=R_{B}^{\theta }(r)\Theta _{B}^{\theta }(\theta ),  \tag{18}
\end{equation}
and similarly for the electric field components $E^{r}$ and $E^{\theta }$:

\begin{equation}
E^{r}(r,\theta )=R_{E}^{r}(r)\Theta _{E}^{r}(\theta ),\quad E^{\theta
}(r,\theta )=R_{E}^{\theta }(r)\Theta _{E}^{\theta }(\theta ).  \tag{19}
\end{equation}
Substituting from equation (18) into equation (10) and simplifying we obtain

\begin{equation}
\frac{1}{\Theta _{B}^{\theta }}\frac{d\Theta _{B}^{\theta }}{d\theta }+\cot
\theta =-(\frac{\Theta _{B}^{r}}{\Theta _{B}^{\theta }})\frac{1}{%
r^{2}u^{t}R_{B}^{\theta }}\frac{d(r^{2}u^{t}R_{B}^{r})}{dr}.  \tag{20}
\end{equation}
For separation we require that $\Theta _{B}^{r}=\Theta _{B}^{\theta }=\Theta
_{B}$ and obtain

\begin{equation}
\frac{1}{\Theta _{B}}\frac{d\Theta _{B}}{d\theta }+\cot \theta =k_{1,} 
\tag{21}
\end{equation}
and

\begin{equation}
-\frac{1}{r^{2}u^{t}R_{B}^{\theta }}\frac{d(r^{2}u^{t}R_{B}^{r})}{dr}=k_{1},
\tag{22}
\end{equation}
where $k_{1}$ is the separation constant. Solving for $\Theta _{B}$ the
equation (21) we obtain $A_{1}\exp k_{1}\theta (\csc \theta )$. Notice that
because of the exponential function the field at coincident points $\theta
=\pi /2$ and $\pi /2+2\pi $ is not identical. We therefore require single
valued solution $k_{1}=0$. The solutions to equations (21) and (22) are now
given by:

\begin{equation}
\Theta _{B}=\frac{A_{1}}{\sin \theta },  \tag{23}
\end{equation}
and

\begin{equation}
R_{B}^{r}=\frac{A_{2}}{r^{2}u^{t}},  \tag{24}
\end{equation}
where $A_{1}$ and $A_{2}$ are constants. Therefore we have

\begin{equation}
B^{r}(r,\theta )=\frac{A}{r^{2}u^{t}\sin \theta },\quad B^{\theta }(r,\theta
)=R_{B}^{\theta }(r)\frac{A_{1}}{\sin \theta },  \tag{25}
\end{equation}
where $A=A_{1}A_{2}$. And similarly from equation (13) and (18) we obtain

\begin{equation}
\Theta _{E}=\frac{C_{1}}{\sin \theta },\quad R_{E}^{r}=\frac{C_{2}}{%
r^{2}u^{t}},  \tag{26}
\end{equation}
and hence

\begin{equation}
E^{r}(r,\theta )=\frac{C}{r^{2}u^{t}\sin \theta },\quad E^{\theta }(r,\theta
)=R_{E}^{\theta }(r)\frac{C_{1}}{\sin \theta },  \tag{27}
\end{equation}
where $C=C_{1}C_{2}$ and $\Theta _{E}=\Theta _{E}^{r}=\Theta _{E}^{\theta }$.

Now to determine the functions $R_{E}^{\theta }$ we substitute from (25) and
(27) into equation (13). After some simplification we obtain in a separated
form:

\begin{equation}
\frac{d(u_{t}r^{2}R_{E}^{\theta })}{dr}=\frac{\sin \theta }{C_{1}}(A\frac{%
d\omega }{dr}-\frac{C}{r^{2}}\frac{\cot \theta }{\sin \theta }).  \tag{28}
\end{equation}
As before we have

\begin{equation}
\frac{d(u_{t}r^{2}R_{E}^{\theta })}{dr}=k_{2},  \tag{29}
\end{equation}
and

\begin{equation}
A\frac{d\omega }{dr}\sin \theta -\frac{C}{r^{2}\sin \theta }\cos \theta
-k_{2}C_{1}=0,  \tag{30}
\end{equation}
where $k_{2}$ is a constant. Since (30) holds for every value of $\theta $,
on comparing coefficients of $\sin \theta $, $\cos \theta $, and unity we
obtain $C=0=k_{2}$ and $A=0$. Here are now two possibilities for non-trivial
solutions depending on either $A_{1}\neq 0$ or $A_{2}\neq 0$ (i.e. $A_{1}=0$%
).

For $A_{1}\neq 0$ equation (29) gives

\begin{equation}
R_{E}^{\theta }=\frac{A_{3}}{u_{t}r^{2}}.  \tag{31}
\end{equation}
where $A_{3}$ is a constant. From equations (25), (27) and (17) we obtain in
a similar manner

\begin{equation}
R_{B}^{\theta }=\frac{C_{3}}{u_{t}r^{2}}.  \tag{32}
\end{equation}
Hence

\begin{equation}
B^{r}(r,\theta )=0,\quad B^{\theta }(r,\theta )=\frac{A_{1}A_{3}}{%
u_{t}r^{2}\sin \theta },  \tag{33}
\end{equation}
and

\begin{equation}
E^{r}(r,\theta )=0,\quad E^{\theta }(r,\theta )=\frac{C_{1}C_{3}}{%
u_{t}r^{2}\sin \theta }.  \tag{34}
\end{equation}
Also from equations (11), (12) and (15), (16) we obtain

\begin{equation}
B^{\varphi }=\frac{A_{4}}{u_{t}r^{2}\sin ^{2}\theta },  \tag{35}
\end{equation}

\begin{equation}
E^{\varphi }=\frac{C_{4}}{u_{t}r^{2}\sin ^{2}\theta }.  \tag{36}
\end{equation}

For $A_{1}=0$ (and similarly for $C_{1}=0$), there is no poloidal component
to the magnetic (electric) field \ and the only non-vanishing components to
the electromagnetic field are given by equations (35) and (36). Clearly in
both cases the electromagnetic field is independent of $\omega (r)$.

In Figures (1) and (2) we give plots for the magnetic (electric) field
components $B^{\theta }$ ($E^{\theta }$) and $B^{\varphi }$ ($E^{\varphi }$)
for $A_{1}A_{3}=1$ ($=C_{1}C_{3}$) and $A_{4}=1$ ($=C_{4}).$

\section{Trajectory of a Charged Particle}

According to the general theory of relativity the trajectory of a particle
in the field of a massive object is determined by the geodesic equation:

\begin{equation}
\frac{d^{2}x^{\alpha }}{ds^{2}}+\Gamma _{\beta \gamma }^{\alpha }\frac{%
dx^{\beta }}{ds}\frac{dx^{\gamma }}{ds}=0;  \tag{37}
\end{equation}
where $\Gamma _{\beta \gamma }^{\alpha }$ is the Christoffel symbol. To
linearize the geodesic equation we assume that the metric tensor can be
expressed as: 
\begin{equation}
g_{\alpha \beta }=\eta _{\alpha \beta }+h_{\alpha \beta }  \tag{38}
\end{equation}
where $\eta _{\alpha \beta }$ $=diag(-1,-1,-1,1)$ is the Minkowski metric
tensor and $h_{\alpha \beta }$ is a small perturbation to the spacetime
metric such that $h_{\alpha \beta }\ll 1$. We can take time $t$ for the
affine parameter $s$. Further requiring $v\mathbf{\equiv \mid }d\mathbf{r}%
/dt\mid \ll 1$ for a `slow' moving particle we obtain$^{15}$ 
\begin{equation}
\frac{d^{2}\mathbf{r}}{dt^{2}}=(\mathbf{G}+\mathbf{v}\times \mathbf{H)} 
\tag{39}
\end{equation}
where

\begin{equation}
\mathbf{G=-\nabla }\varphi ,\quad \mathbf{H=\nabla \times }4\mathbf{a} 
\tag{40}
\end{equation}
and

\begin{equation}
\varphi =-\iiint \frac{\rho }{r}dV,\quad \mathbf{a=}\iiint \frac{\rho 
\mathbf{v}}{r}dV  \tag{41}
\end{equation}
$\rho $ being the mass density and $V$ is the volume of the gravitational
source. This approximation, due to its formal analogy to the classical
electromagnetic theory, is referred to as the gravitoelectromagnetic (GEM)
approximation to the general theory of relativity (see reference [16] and
[17] for different interpretations). Thus in expression (40) $\mathbf{G}$ is
called the gravitoelectric (GE) force per unit mass whereas the term $m%
\mathbf{v\times H}$ is called the gravitomagnetic (GM) force for a unit
mass. In particular the GM force, being independent of the choice of a
particular coordinate system used, is general relativistically significant.

In the slow rotation approximation the deformation to the star due to the
effects of rotation are small compared to the \ radially attractive GE
force; hence we take the star to be a slowly rotating sphere of homogeneous
mass density. Then using GEM analogy, we obtain:

\begin{equation}
\mathbf{G}=-M\hat{\mathbf{r}}/r^{2},  \tag{42}
\end{equation}
and 
\begin{equation}
\mathbf{H}=-\frac{12}{5}MR^{2}(\mathbf{\Omega .r}\frac{\mathbf{r}}{r^{5}}-%
\frac{1}{3}\frac{\mathbf{\Omega }}{r^{3}}),  \tag{43}
\end{equation}
where $M$ is the mass, $R$ is the radius and $\mathbf{\Omega }$ is the
angular velocity of the star. Thus in the field of a slowly rotating compact
star, the net force acting on a particle of mass $m$ and charge $q$ is the
sum of the gravitational and the electromagnetic forces:

\begin{equation}
m\frac{d^{2}\mathbf{r}}{dt^{2}}=m(\mathbf{G}+\mathbf{v}\times \mathbf{H)+}q%
\mathbf{(E+v\times B).}  \tag{44}
\end{equation}

To specify the physical situation we study the motion of the charged
particle in a Cartesian coordinate system $(x,y,z)$ and take $\mathbf{H}$ to
be along the positive $z$-axis. Furthermore, since now the effective
component of the GM force lies in the $XY$-plane, we assume that the $XY$%
-plane is the equatorial plane of the star and the particle's motion is
confined to this plane only.

Then for the case of non-vanishing poloidal magnetic field we transform
expressions (33)-(36) and (42)-(43) into the Cartesian coordinates, and
obtain the equations of motion in the equatorial plane (i. e. for $\theta
=\pi /2$):

\begin{eqnarray}
\frac{d^{2}x}{dt^{2}} &=&-\frac{M}{x^{2}+y^{2}}+\frac{yE_{0}}{\sqrt{%
(x^{2}+y^{2})^{2}-2M(x^{2}+y^{2})^{3/2}}}+  \nonumber \\
&&(\frac{\mu }{(x^{2}+y^{2})^{3/2}}-\frac{B_{0}}{\sqrt{x^{2}+y^{2}-2M\sqrt{%
x^{2}+y^{2}}}})\frac{dy}{dt}\mathbf{,}  \TCItag{45}
\end{eqnarray}
\begin{eqnarray}
\frac{d^{2}y}{dt^{2}} &=&-\frac{M}{x^{2}+y^{2}}-\frac{yE_{0}}{\sqrt{%
(x^{2}+y^{2})^{2}-2M(x^{2}+y^{2})^{3/2}}}-  \nonumber \\
&&(\frac{\mu }{(x^{2}+y^{2})^{3/2}}-\frac{B_{0}}{\sqrt{x^{2}+y^{2}-2M\sqrt{%
x^{2}+y^{2}}}})\frac{dx}{dt}\mathbf{,}  \TCItag{46}
\end{eqnarray}
where $B_{0}=qA_{1}A_{3}/m$ , $E_{0}=qC_{4}/m$ and $\mu =4MR^{2}\Omega /5$.
Notice that here the case of vanishing poloidal magnetic field is obtained
for $B_{0}=0$ whereas requiring $C_{1}=0$ does not alter the equations of
motion (45) and (46). We numerically solve the equations of motion (45) and
(46) for the initial conditions $x(0)=10=y(0)$ and $dx/dt\mid
_{t=0}=0=dy/dt\mid _{t=0}$, and plot (in Figure (3) to (6)) the solutions
for various values of the parameters $M$, $\mu $, $E_{0}$, and $B_{0}$.

\section{Discussion and Conclusions}

We have considered the dynamics of a charged particle in a highly,
conducting plasma surrounding a slowly rotating compact gravitational
source. The effects of background spacetime have been discussed, both
directly on the particle's motion and indirectly via an axially symmetric
electromagnetic field around the star.

We found (equations (33) to (36)) that when slow rotation is assumed for the
compact source the spacetime dragging does not modify or reduce the
electromagnetic field of the star. Hence the field remains, in this
approximation, the same as for the Schwarzchild spacetime. However the
motion of the particle is directly influenced via gravitational field of the
star.

Our numerical analysis of the effects of gravitational and electromagnetic
field shows (Fig. (3) \ to (6)) that whereas the electromagnetic effects
dominate the overall dynamics of the charged particle at sufficiently large
distances, the gravitational field also becomes important closer to the
surface of the star especially if the star is rotating fast enough. Firstly
we observe that the effects of the electric field is to lift a charge
particle from close to the surface of the star (Fig. (3)) and make them move
along the magnetic field lines. Here are now two possibilities: one, if the
magnitude of the magnetic field is larger than the magnitude of the electric
field the particle following a helical trajectory falls into the star (case
(a), (b), and (c) in Fig. (4)); two, if the magnetic field strength is
weaker or is even comparable to the electric field strength, then the
electrical effects become dominant and the particle escapes from falling
into the star (case (d), and (e) in Fig.(4)).

On the other hand the effects of gravitational field can be regarded as
opposing that of the electromagnetic field. In Figure (5) we notice that\
the GE force attracts the particle towards the star's surface against the
electric field. Furthermore it is clear from Figure (6) that the gyroradius
of the helical trajectory of the particle is increased due to the GM force;
hence the GM force weakens the effects of the magnetic field.

Summing up these observations we note that the accretion of a charged
particle in vicinity of a compact star is mainly due to the GE and magnetic
effects. These effects are especially dominant at a sufficiently large
distances from the star. However \textit{close} to the star the particle
trajectory is also effected by the GM force as well as electric field which
cause accerted charged particles to follow open lines of force. For a
sufficiently fast rotating star and an electric field comparable to the
magnetic field the above mechanism (GM + electric) may contribute to
opposing the in-fall of the plasma surrounding the compact star.

\textbf{Acknowledgments}

Useful comments of Dr. H. Saleem and Dr. D. V. Ahluwalia are gratefully
acknowledged.

\bigskip

\textbf{Figure Captions}

1) Plot for the magnetic (electric) field component $B^{\theta }$ ($%
E^{\theta }$) as a function of the radial distance $r$ and polar angle $%
\theta $ for $A_{1}A_{3}=1(=C_{1}C_{3})$.

2) Plot for the magnetic (electric) field component $B^{\varphi }$ ($%
E^{\varphi }$) as a function of the radial distance $r$ and polar angle $%
\theta $ for $A_{4}=1(=C_{4})$.

3) Trajectory of the charged particle for electric force of varying strength
with $E_{0}=0.1,1,5,10,20$, and $M=1$, $\mu =0.1$, $B_{0}=10$ in
gravitational units.

4) Trajectory of the charged particle for magnetic force of varying strength
with $B_{0}=0,10,30,50,70$, and $M=1$, $\mu =0.1$, $E_{0}=1$ in
gravitational units.

5) Trajectory of the charged particle for GE force of varying strength with $%
M=1,2,3,5,10$, and $\mu =0.1$, $E_{0}=1$, $B_{0}=10$ in gravitational units.

6) Trajectory of the charged particle for GM force of varying strength with $%
\mu =0.1,0.2,0.3,0.5,0.8$, and $M=1$, $E_{0}=1$, $B_{0}=20$ in gravitational
units.

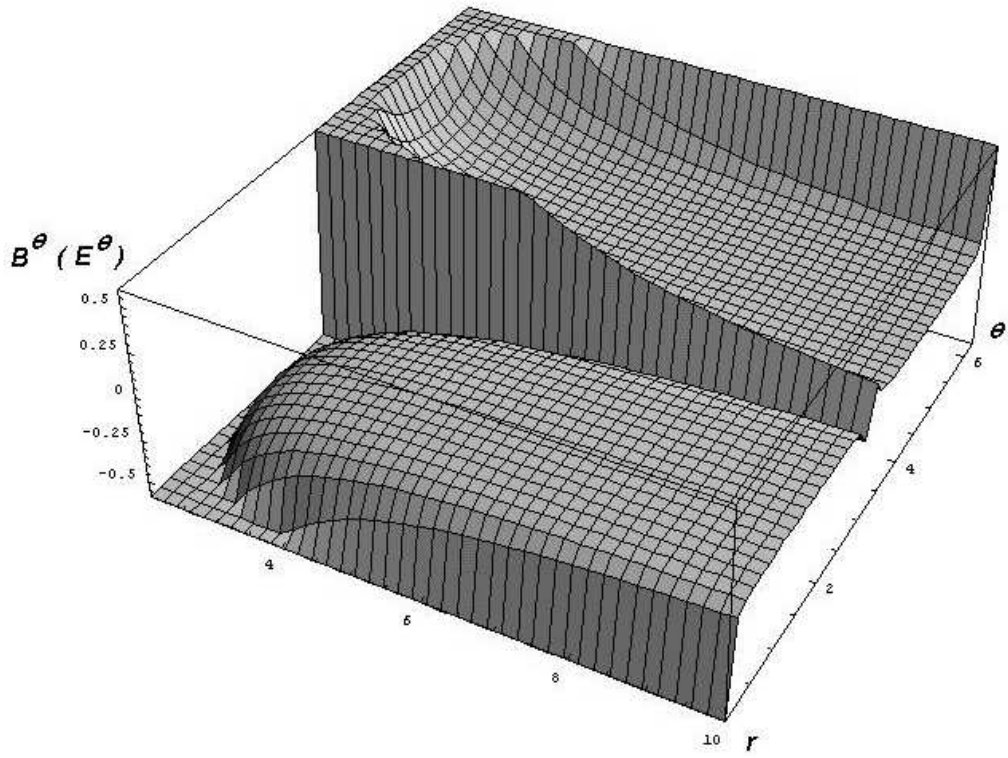
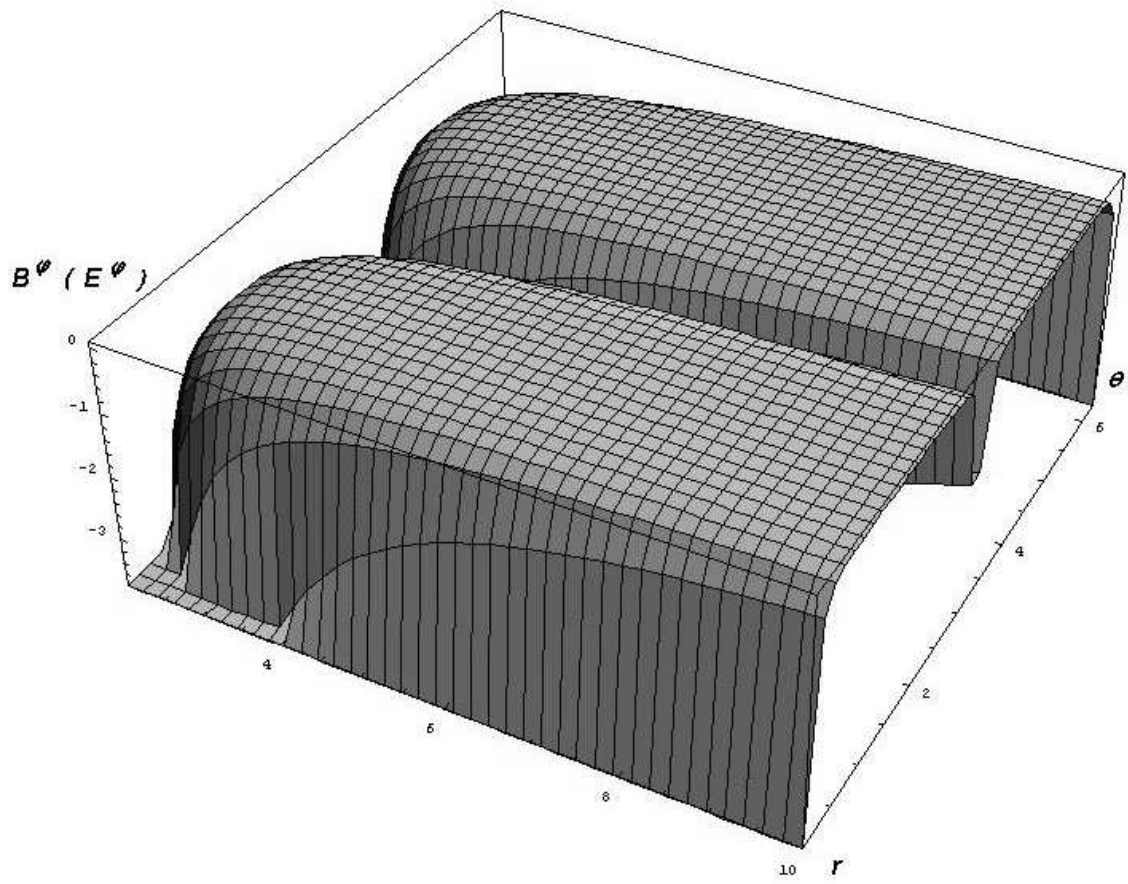
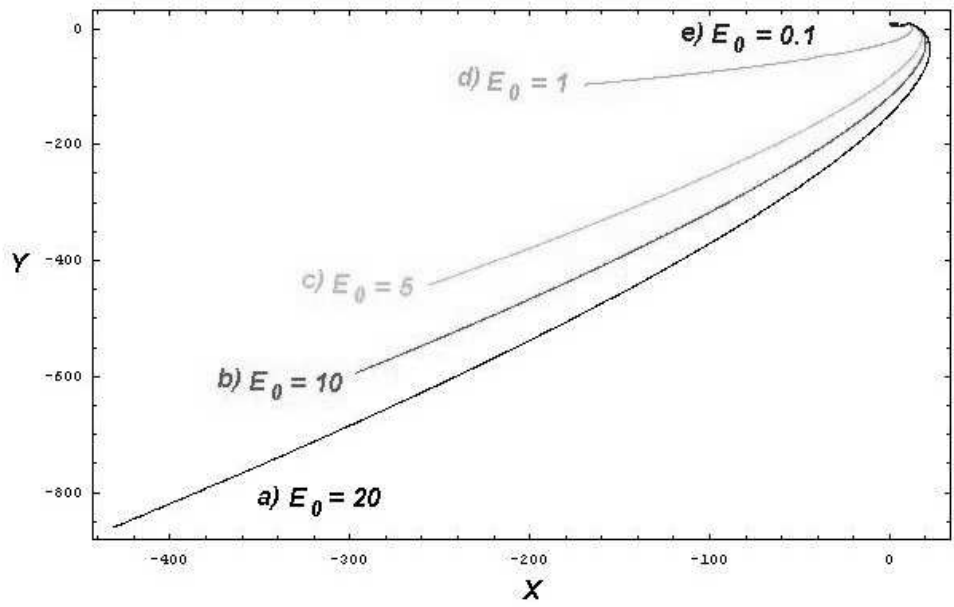
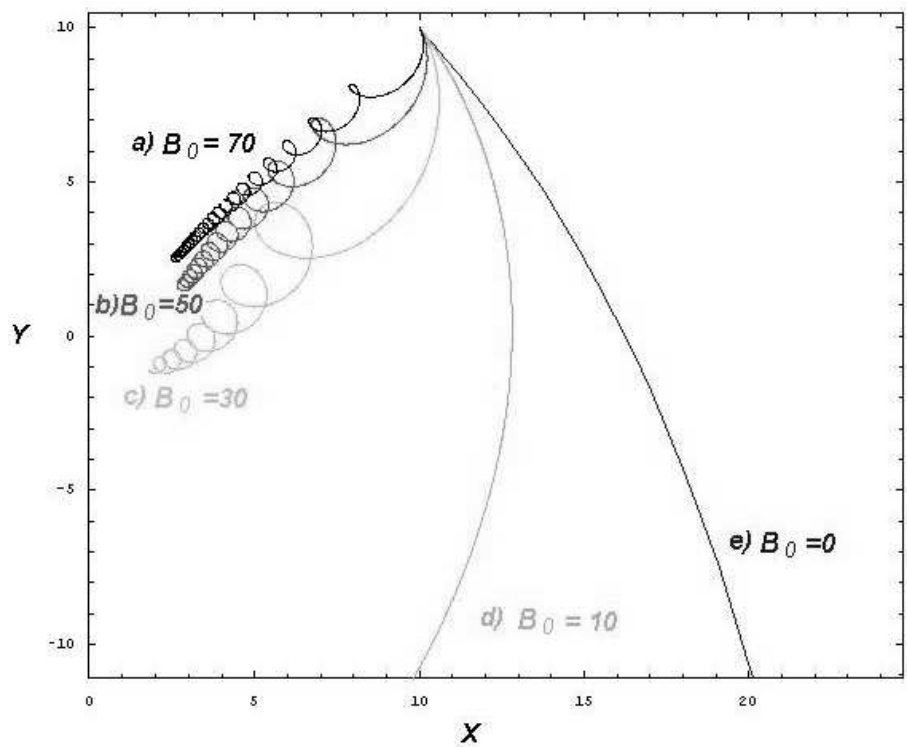
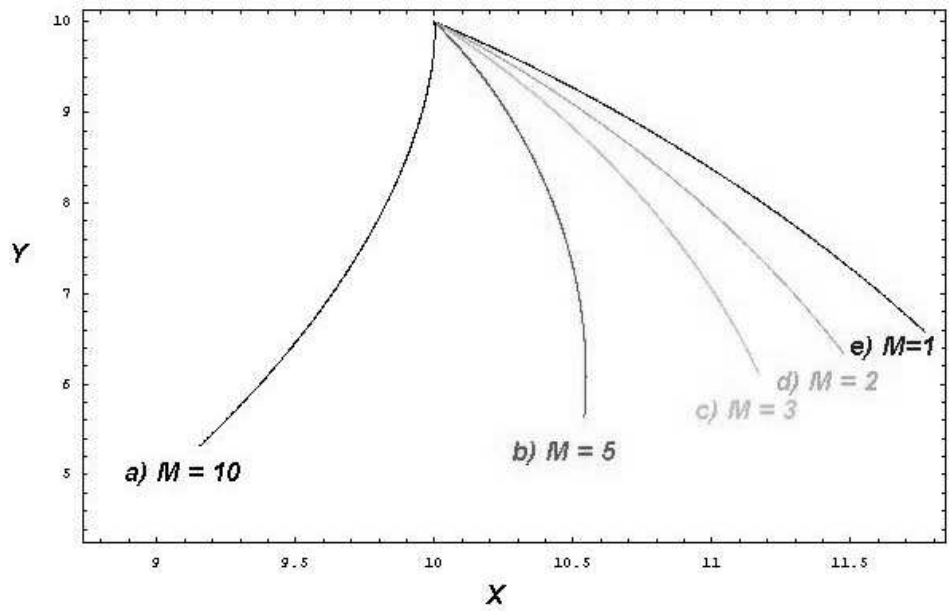
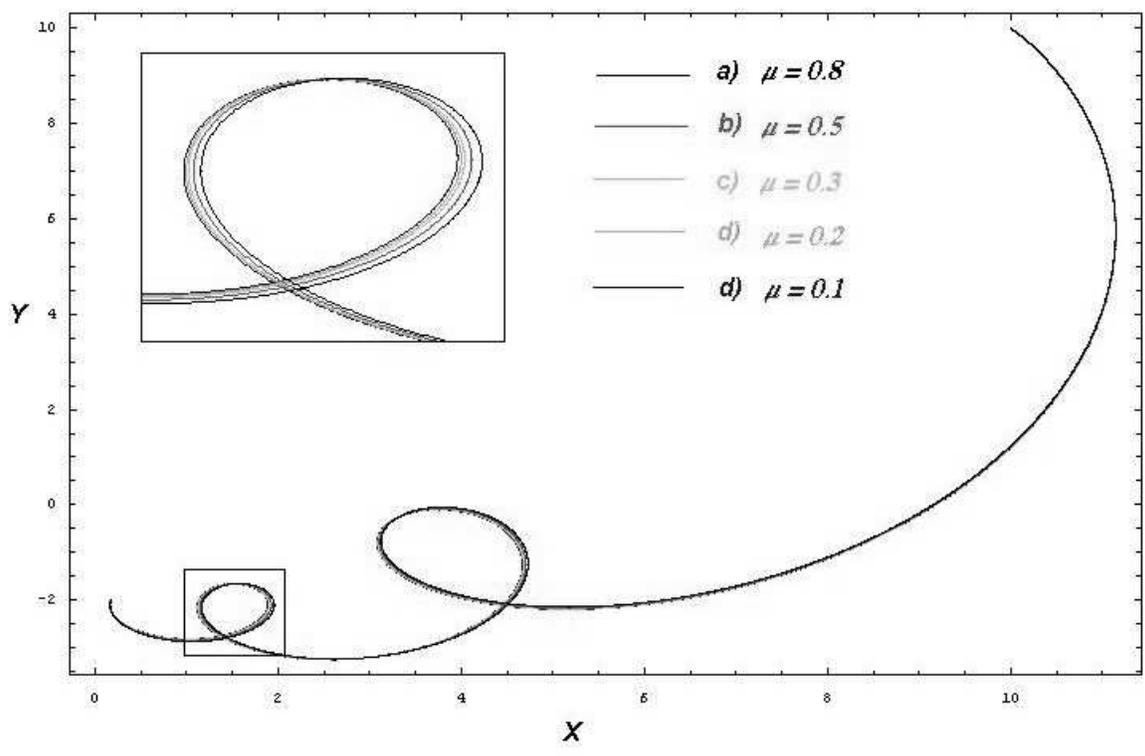
\end{document}